\documentclass[10pt]{article}

\usepackage{amsmath}
\usepackage{amssymb}

\usepackage{graphicx}

\usepackage{cite}

\usepackage{color}

\usepackage{url}

\usepackage[labelfont=bf,labelsep=period,justification=raggedright]{caption}

\makeatletter
\renewcommand{\@biblabel}[1]{\quad#1.}
\makeatother

\date{}

\begin{document}

% Title must be 150 characters or less
\begin{flushleft}
{\Large
\textbf{Exploring the complex pattern of information spreading in online blog communities}
}
% Insert Author names, affiliations and corresponding author email.
\\
Sen Pei$^{1,2 *}$,
Lev Muchnik$^{3}$,
Shaoting Tang$^{1}$,
Zhiming Zheng$^{1}$,
Hern\'an A. Makse$^{2}$
\\
$^{1}$ Laboratory of Mathematics, Informatics and Behavioral Semantics, and School of Mathematics and Systems Science, Beihang University, Beijing, China
\\
$^{2}$ Levich Institute and Physics Department, City College of New York, New York, USA
\\
$^{3}$ School of Business Administration, The Hebrew University of Jerusalem, Israel
\\
$^{*}$ Corresponding author
\\
 E-mail: peisen@buaa.edu.cn (SP)
\end{flushleft}

% Please keep the abstract between 250 and 300 words
\section*{Abstract}

Information spreading in online social communities has attracted tremendous attention due to its utmost practical values in applications. Despite that several individual-level diffusion data have been investigated, we still lack the detailed understanding of the spreading pattern of information. Here, by comparing information flows and social links in a blog community, we find that the diffusion processes are induced by three different spreading mechanisms: social spreading, self-promotion and broadcast. Although numerous previous studies have employed epidemic spreading models to simulate information diffusion, we observe that such models fail to reproduce the realistic diffusion pattern. In respect to users behaviors, strikingly, we find that most users would stick to one specific diffusion mechanism.  Moreover, our observations indicate that the social spreading is not only crucial for the structure of diffusion trees, but also capable of inducing more subsequent individuals to acquire the information. Our findings suggest new directions for modeling of information diffusion in social systems and could inform design of efficient propagation strategies based on users behaviors.

\section*{Introduction}

Information spreading is a key social phenomenon, which due to the prevalence of communication and information technologies in the recent years is becoming increasingly dynamic and powerful in shaping such processes as the adoption of innovations \cite{Rogers1995}, the propagation of news and information \cite{Watts2007,Kleinberg2008,Muchinik2013a}, the success of viral marketing \cite{Watts2007a,Leskovec2007,Van2007,Aral2014} as well as the spread of behaviors and social norms in general \cite{Christakis2009,Centola2010}. With the mainstream adoption of Internet and World Wide Web, several types of user-based information dissemination platforms, such as the blogs sharing communities, online social networks, and microblog services, have become prevalent. In particular, the blogspace has attracted millions of users to discuss or share topics ranging from public concerned affairs to users' personal lives, thus greatly facilitating people to create, capture and disseminate information \cite{Gruhl2004,Java2006,Gallos2012,Guille2013}. Indeed, people's life and social relations have been influenced to a great extent by the information in the blogspaces. As a consequence, understanding the mechanism by which a piece of information diffuses through population is crucial for not only designing efficient promotion strategies but also blocking the pervasion of malicious information. Because of its practical values in real-world applications, information spreading in online communities has attracted much attention across disciplines. In particular, many studies have been addressing to model the information spreading process \cite{Lu2011,Liu2014,Zhu2014}.

Much previous research investigating the information spreading among individuals has been motivated by analogy with the spread of a contagious disease \cite{Hethcote2000,Pei2013,Barrat2008,Kitsak2010,Yan2013}. In these disease-propagation models, a piece of information is assumed to diffuse along social links in the underlying social network from person to person complying with specific rules and may eventually reach a large fraction of the population \cite{Kleinberg2007}. With these assumptions, it has been shown that the social network structure can significantly affect the outcome of a spreading process \cite{Pei2013,Barrat2008,Satorras2001}. Although such epidemic-motivated approaches have led to many profound results, these models are typically based on rather simplified assumptions that may not be representative for information spreading in real circumstances. Recent observational studies tracking actual diffusion processes demonstrate that information and diseases spread differently: apart from the network structure, the contagion of information is also affected by other factors, including the human behavior \cite{Iribarren2009,Funk2010,Muchnik2013}, homophily \cite{Centola2011,Aral2013}, social reinforcement \cite{Centola2010}, etc. With the existence of such factors, the diffusion paths of information in real world are found to be dramatically different from those derived from the classical epidemic models \cite{Goel2012}. In addition, recent empirical studies on information flows in blogspaces have presented evidences that information does not necessarily diffuse via social links \cite{Grabowicz2012,Pei2014,Li2014}, which opposes to the prevailing assumption that information spreading is confined within the underlying social networks. All these observations of real-world diffusion instances indicate that we still lack the systematic understanding of how information diffuses in online social communities.

In this paper, we perform a detailed analysis of information spreading in an online blog community - LiveJournal. Earlier studies of the individual-level diffusion patterns in blogspaces \cite{Gruhl2004,Java2006,Gallos2012,Guille2013} focused on the topological feature of diffusion paths. Here, by analyzing information diffusion trees along with the social network structure, we seek to unveil the diffusion pattern of information spreading in blog communities based on users' behaviors of disseminating information. We find that the spreading instances can be classified into three types: {\it social spreading}, which occurs following social links, {\it self-promotion}, meaning references of earlier posts by the same author, and {\it broadcast}, representing the remaining cases. A typical spread of information in social network can therefore be seen as a combination of these three processes. We explore the topological structure associated with each of the three categories and find that the vast majority of diffusion cases are confined to just first few steps. Through simulations of the susceptible-infected-recovered (SIR) model with real-world infection rate, we find that the SIR model fails to reproduce the realistic social spreading pattern. This distinguishes information diffusion in social networks from the epidemic spreading. In respect of the adoption of specific behavioral pattern by users, it is striking that nearly all the users would stick to an inherent diffusion type. Such persistence in behavior of the individual users can be interpreted as manifestation of their preferences and the role they pursuit in the social network. We proceed by further investigating the dynamical coupling of distinct diffusion mechanisms with a particular focus on social spreading. Although, social spreading is observed less frequently than the other mechanisms, it tends to be associated with larger cascades and can induce more subsequent disseminations. This finding implies that, with the propelling of social spreading, information could reach a much larger fraction of the population, which highlights the essential role of social spreading in information diffusion. From the perspective of human behaviors, our findings reveal a set of complex behavioral patterns exploiting different information spread mechanisms and suggest new directions for modeling information spreading in online social networks.

\section*{Materials and Methods}

In order to explore the information diffusion in LiveJournal.com, we have obtained approximately 56 million posts published during a period of 21 months. The social network constructed by friend lists consists of 9,573,127 users and 188,240,039 social links. Aimed to track the information flow from one user to the other, we analyze the hyperlinks contained in the posts. To be precise, for each post containing hyperlinks to other posts in LJ, we record the following information: ID of the post $P_{d}$, ID of the post's author $I_{d}$, publication time of the post $T_{d}$, ID of the referred post $P_{s}$, ID of the referred post's author $I_{s}$ and the time of the referred post's publication $T_{s}$. In total, we have identified 3,462,504 posts referencing other posts in our dataset. Using this data, we are able to track the information flow as follows: user $I_{d}$ has published a post $P_{d}$ at time $T_{d}$, which contains the hyperlink to post $P_{s}$ published by user $I_{s}$ at time $T_{s}$. Therefore, the information diffuses from the source user $I_{s}$ to the destination user $I_{d}$ explicitly.

The posts data is composed of public posts transmitted by the company hosting the platform via the live stream (http://www.livejournal.com/stats/latest-rss.bml). We do realize that this data, even if classified as public and freely accessible online, may potentially contain private information. Our study is therefore limited to the exploration of the hyperlinks between the posts, rather than analysis of their textual content. Furthermore, we anonymize the dataset by replacing user names with numerical IDs. A complementary data, the social network structure, was obtained by crawling the LiveJournal web site via the publicly available API, FOAF (http://www.livejournal.com/bots/) designed specifically for that purpose. We would like to note that all social network connections are public. Still, we anonymize the network substituting LiveJournal user names with numerical IDs.

% Results and Discussion can be combined.
\section*{Results}
In pursuing to unveil the patterns of information spreading in online social communities, we have collected the complete social network structure as well as the available blog posts of a large-scale online social community - LiveJournal.com (LJ). The details of collected data are explained in Materials and Methods. The LJ platform facilitates maintenance of friend lists that help users track information updated by their peers, thus facilitating information spreading among them. The resulting social network, which consists of nearly 9.6 million users, records the complete information of friend relations in the LJ community. This network composed of social links has been explored in several previous studies and is believed to reliably represent actual social relations in the community \cite{Backstrom2006,Liben2005}. In this work we leverage the LJ lists of followers to construct an undirected graph and use it to track and study propagation of information in that social network. We identify and track the propagating pieces of information by analyzing the content of posts. In LJ community, posts typically refer (by heperlinking) the source of the information which they retransmit, discuss or refer to. By extracting these hyperlinks and matching them to earlier posts, we can directly track the information flow from one user to the other (details in Materials and Methods). This inference technique confines the information flow to the LJ community and excludes the diffusion content coming from external sources, such as news channels and newspapers.

By analyzing the relationship between the source user $I_{s}$ and destination user $I_{d}$ in each spreading instance, the spreading pattern can be categorized into three groups. We attribute the content spreading between the users explicitly connected in social network ($I_{s}$ and $I_{d}$ are linked) to social spreading by reasoning that such propagation is induced by the underlying social network. We classify the posts citing same user's earlier posts ($I_{s}=I_{d}$) as self-promotion. The third, broadcast news propagation pattern, encompasses the posts citing earlier publications of remote (in terms of the underlying social network) users. The broadcast pattern typically relies on search, promotion and the LJ recommender engine to discover the content from remote users. In the collected data, these three types of information diffusion, i.e., the social spreading, self-promotion, and broadcast, make up $26.8\%$, $31.14\%$, and $42.06\%$ of the total propagation respectively. This suggests that, contrary to the common belief that information flow is confined to the social network, the real information propagation in online social community exhibits a complex pattern in which the mentioned three diffusion patterns coexist and entwine with each other.

Compared with other online social systems, such as Twitter where the social influence is more than $70\%$ \cite{Myers2012}, the composition of diffusion links of social spreading is only $26.8\%$. The difference may come from the features of LJ system. LiveJournal is a social networking service where users can keep a blog, journal or diary. It not only has the typical social function of ``friend list'', but also maintains large numbers of communities, which are collective blogs in which different users can post messages. Users who are interested in a particular subject can find or create a community for this subject. Notice that users in the same community are not necessarily connected by friend links. Therefore, even though users repost information within one community, these diffusion links will still be classified as broadcast. Our data indicates that over $40\%$ diffusion processes are attributed to such broadcast. And the proportion of social spreading is therefore squeezed to $26.8\%$.

The named diffusion patterns represent distinct propagation mechanisms. Even when social spreading is considered, the information diffusion processes differ fundamentally from the classical viral contagion. The information recipient exposed to the content is the one who decides whether to adopt the behavior and share the information among her own local social network. The authors cannot control such kind of passive diffusion. As has been pointed recently \cite{Goel2012}, the behavior adoption is not completely consistent with the classical epidemic spreading processes, in which the viruses are capable of self-replicating and propagating themselves proactively. Unlike social spreading, the self-promotion mechanism relies on the dedication of the authors to repeatedly promote their own content, increasing exposure and the probability of consequent sharing (i.e. see complex contagion \cite{Centola2007}). In contrast to social spreading and self-promotion, broadcast is more likely to be caused by the mechanism of marketing or mass media. These arguments show that, although the viral disease-like diffusion has been intensively explored in previous literature \cite{Pastor2014}, the dynamics of such coupled information spreading processes is still remained to be explored.

\subsection*{Construction of diffusion trees}

\begin{figure}[ht]
\begin{center}
\includegraphics[width=0.8\columnwidth]{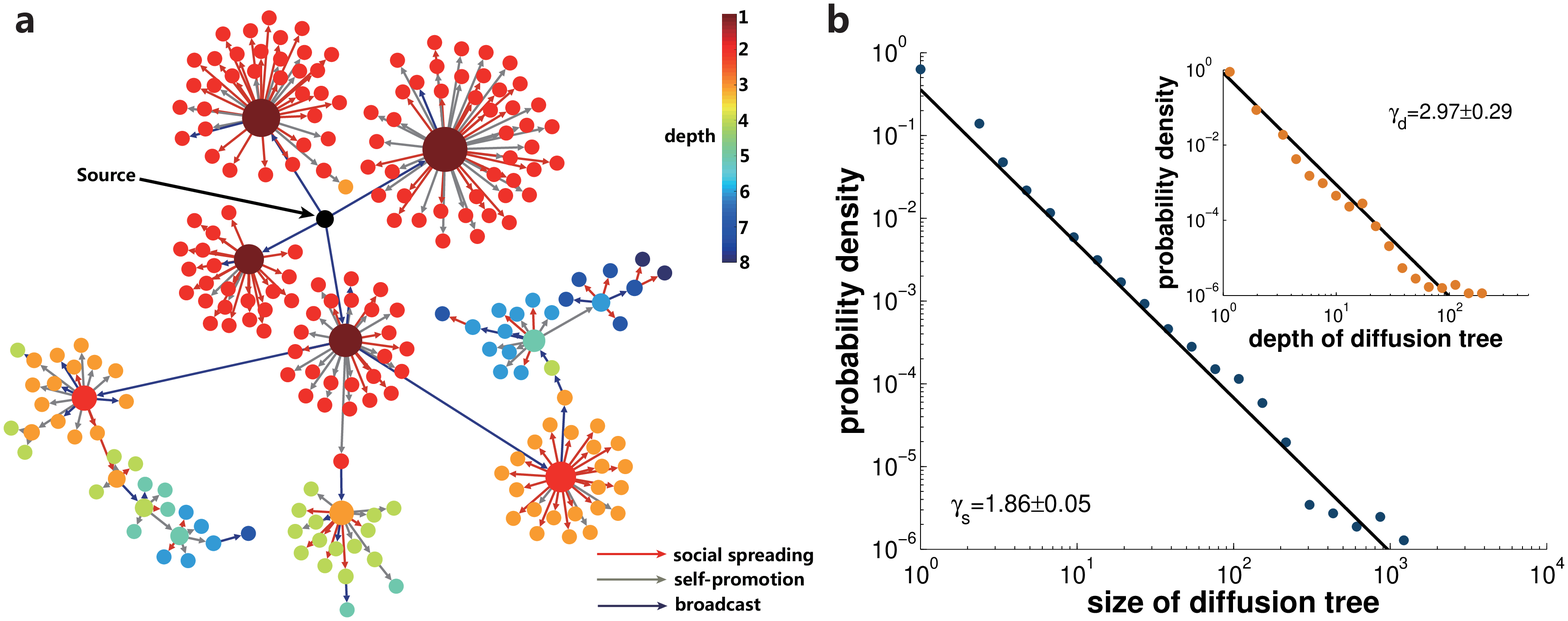}
\end{center}
\caption{{\bf The diffusion trees in LJ community.} {\bf a}, a real instance of information diffusion. An illustration of a diffusion tree containing 227 nodes which reaches the depth of 8. Each node represents a post published in LJ community, whereas each link stands for a spreading instance. The node color indicates the depth of a node in the diffusion tree. The size of a node is proportional to the number of its children. The links of social spreading, self-promotion, and broadcast are represented by the colors of red, grey, and blue respectively. {\bf b} shows the probability distributions of diffusion trees' size and depth. Both the tree size and depth exhibit approximately pow-law distributions. The power-law exponents for tree size and depth are $\gamma_{s}=1.86\pm0.05$ and $\gamma_{d}=2.97\pm0.29$ respectively. The straight lines represent the maximum likelihood fitting of the data points.} \label{diffusiontree}
\end{figure}

Aiming to acquire the individual-level details of information spreading, we have reconstructed the exact diffusion tree for each source information. More specifically, we start by sorting the posts by their publication time $T_{s}$ so that the cited posts are guaranteed to appear chronologically in the record list. Then we identify the source information which has only been referred by other posts but contains no hyperlinks to other posts. For each of these source information, we create a search queue to find the subsequent diffusion in a Breadth-First-Search (BFS) fashion: push the IDs of the posts citing the source information into the search queue; then pop out the front element of the queue, and push into the queue the IDs of the posts that cited the popped-out post but are still not visited in previous steps; this process is applied recursively until the search queue becomes empty. We define the diffusion depth to be the number of layers the information spreads out from the origin. And the size of a diffusion tree is defined as the number of posts found in our search. In total, we have reconstructed 880,195 information diffusion trees in LJ community. Figure \ref{diffusiontree}a presents one real diffusion tree of depth 8 and containing 227 nodes. Figure \ref{diffusiontree}b shows the distributions of the size and depth of the observed diffusion trees. These two quantities exhibit power-law distributions with exponents $\gamma$, characterized by  $\gamma_{s}=1.86\pm0.05$ and $\gamma_{d}=2.97\pm0.29$ correspondingly (the values of $\gamma$ are obtained using the maximum likelihood method \cite{Clauset2009}).

The power-law distribution of the diffusion tree size and depth implies that the majority ($63.0\%$) of the posts get to be cited only once. In the posts that acquire numerous citations, diffusion process is fueled by three distinct mechanisms, as can be seen from Fig. \ref{diffusiontree}a. In particular, the links representing social spreading, self-promotion, and broadcast are marked with different colors. It is the complex dynamical interaction between these three kinds of diffusion mechanisms that results in, perhaps infrequent, but significant cases of information propagation in social network. We continue by detailed exploration of each of the three diffusion patterns and the interaction between them.

\subsection*{Social spreading}

In this section, we examine how information propagates via the underlying social network. Much of the current research perceives viral spread of information in full analogy with the spread of contagious diseases \cite{Watts2007a,Leskovec2007}. This has led to the wide adoption and extensive use of the classical disease models like susceptible-infectious-recovered (SIR) and susceptible-infectious-susceptible (SIS) in information propagation studies \cite{Hethcote2000,Pei2013,Barrat2008,Kitsak2010,Satorras2001}.

\begin{figure}[ht]
\begin{center}
\includegraphics[width=0.8\columnwidth]{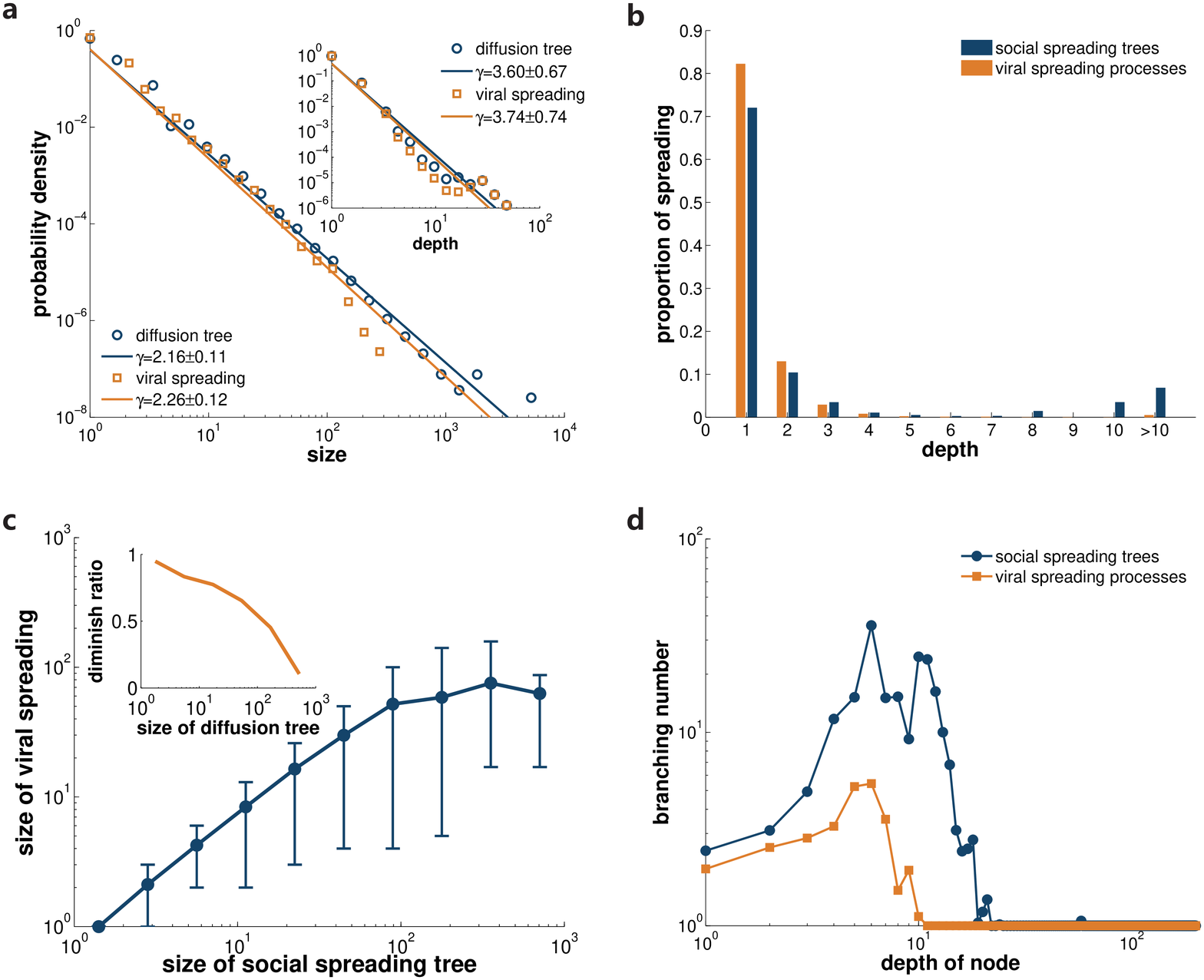}
\end{center}
\caption{{\bf Analysis of social spreading.} {\bf a} shows the probability distributions of the size of diffusion trees and viral spreading processes. The inset displays the distributions of spreading depth for both cases. The straight lines are fitted with the maximum likelihood method. In {\bf b}, we present the proportion of diffusion instances in spreading processes with a given depth. The relation between the size of viral spreading and diffusion trees is displayed in {\bf c}. Error bars indicate the $10\%$ and $90\%$ percentiles. The inset presents the diminishing ratio when mapping the diffusion trees to viral spreading. In {\bf d}, we classify the nodes according to their depth in spreading processes and display their average branching number. } \label{socialspread}
\end{figure}

As we focus on {\it social contagion} in detail, we eliminate the other types of information propagation links and obtain 363,115 distinct diffusion trees entirely composed of social spreading links. The distribution of tree size still follows a power-law shape, although with a larger exponent of $\gamma=2.16\pm0.11$ (Fig. \ref{socialspread}a). Even though, exceptionally deep social diffusion trees do indeed occur, they are too rare to play significant role in information spread processes. We find that over $85\%$ of social spreading is attributed to cascades that do not exceed the depth of 3 (Fig. \ref{diffusiontree}b). Therefore, in online blog communities, the majority of social spreading occurs via small and shallow information cascades. Our observation is in accordance with previous findings in other online communities \cite{Goel2012}. Recall that for epidemic spreading, infections over multiple generations are responsible for most of the contagion. This differs fundamentally from our observation in information spreading. Such difference motivates us to further explore the topological structure of diffusion trees. To delve into this issue, we examine the branching number (number of children) of each post in diffusion trees. More specifically, we identify posts' depth in the diffusion and then display the average branching number for each depth in Fig. \ref{socialspread}d. The average branching number of nodes in the first few generations varies above 1, while for posts located at the depth of more than 20, the branching number is almost 1. As a consequence, for extremely deep diffusions, most of the posts are shared within the first few generations.

Now we turn to explore the relationship between the spreading dynamics and underlying social networks. First, we map the diffusion trees in which nodes represent posts onto the underlying social network in which nodes are individuals. For simplicity, we will refer the spreading among users following social links as viral spreading in our discussion. To avoid unnecessary complexity we represent repeated referrals to the same post by the same user with a single link corresponding to the earliest citation. The obtained viral spreading is therefore represented by a directed graph rather than a tree. Furthermore, numerous appearances of the same user in the diffusion tree translate to a single node in the viral spreading graph, resulting in substantially smaller constructs. Figure \ref{socialspread}c shows the relation between the viral spreading size and diffusion trees' size. The diminishing ratio, which is defined as the ratio between the size of viral spreading and corresponding diffusion trees, decreases significantly as the tree size grows. This means, for larger diffusion trees, there will be more users repost same information repeatedly. Consequently, larger diffusion trees may not necessarily reach larger population. The distribution of viral spreading size is still a power-law, with an exponent $\gamma=2.26\pm0.12$, while the distribution of viral spreading depth remains almost the same as that of diffusion trees (see Fig. \ref{socialspread}a).

The representation of viral spreading graphs helps confirming that most of the diffusion instances are indeed confined to very few steps. In agreement with our earlier observation in diffusion trees, we find that over $95\%$ of spreading instances occur in one to three layers, as shown in Fig. \ref{socialspread}b. If we classify the nodes according to their depth in the viral spreading, we find that for the nodes with depth less than 10, the average branching number remains above 1 (Fig. \ref{socialspread}d). On the contrary, all the nodes deeper than 10 generations have the branching number of 1. This phenomenon indicates that most of the viral spreading occurs in the first few layers.

\begin{figure}[ht]
\begin{center}
\includegraphics[width=1\columnwidth]{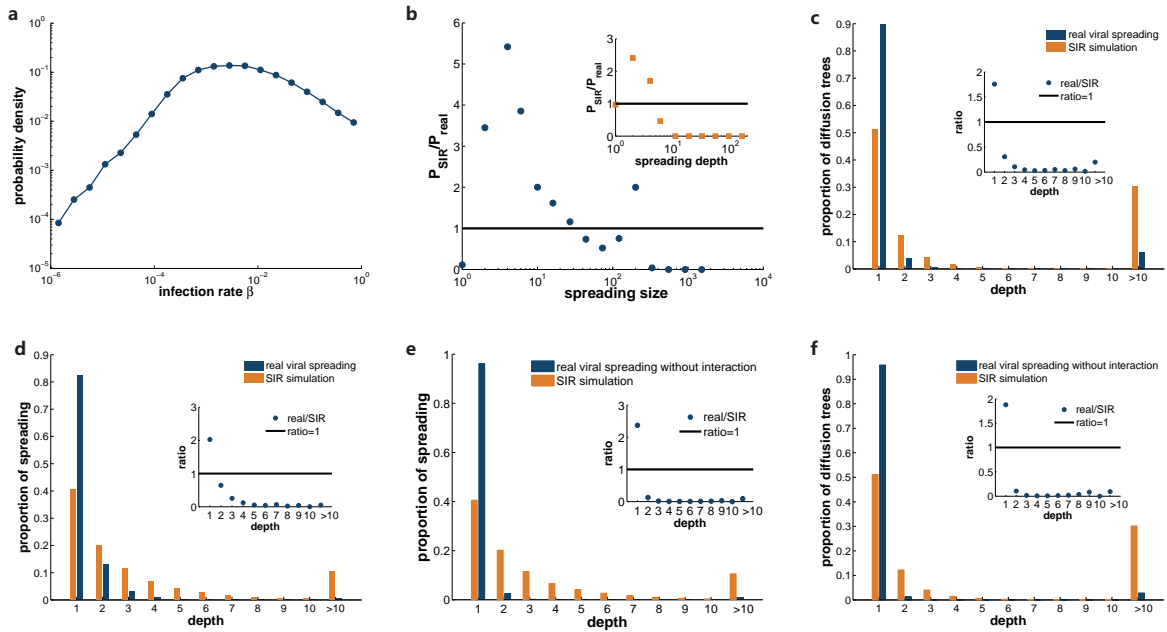}
\end{center}
\caption{{\bf SIR modeling with users' infection rate cannot reproduce the realistic viral spreading pattern.} {\bf a}, distribution of the real-world infection rate for each individual $\beta$ calculated from viral spreading instances. We display the ratio between the size distribution of SIR simulations and real viral spreading in {\bf b}. Inset shows the ratio of depth distribution. In {\bf c}, we present the proportion of diffusion trees with a given depth for both SIR simulations and real viral spreading, and show the ratio between real cases and SIR modeling in the inset. {\bf d} presents the proportion of spreading instances for diffusion with a specific depth for both cases. The inset shows the ratio between real viral spreading and simulations. In {\bf e} and {\bf f}, we perform same analyses for viral spreading without interactions with other diffusion types. } \label{SIR}
\end{figure}

To better explore the difference between the pattern of information and epidemics spreading, we perform extensive SIR simulations with real-world infection rate. In SIR model, infected nodes infect their susceptible neighbors with probability $\beta$ and then enter the recovered state with probability $\lambda$, where they become immunized and cannot be infected again \cite{Kitsak2010,Satorras2001}. In our study, we set $\lambda=1$. For users involved in viral spreading, we define their infection rate $\beta$ as the fraction of neighbors who repost their information. If a user participates in more than one spreading instance, we just adopt the average value of individual infection rate. Unlike the common assumption that individuals have same infection rate in previous studies, the distribution of realistic infection rate is quite heterogeneous (Fig. \ref{SIR}a). Starting from the same spreading sources as empirical viral spreading, we conduct $100$ SIR realizations for each source with realistic infection rate. We take the average size and depth of these realizations as the outcomes of SIR simulations. In Fig. \ref{SIR}b, we display the ratio of size probability density between SIR simulations and real viral spreading $P_{SIR}/P_{real}$. SIR model overestimates the number of moderate diffusion, but underestimates the number of size 1 and extremely large diffusion. For spreading depth, SIR model also bias to diffusion with medium depth (see inset of Fig. \ref{SIR}b). Since SIR modeling and real-world viral spreading have same information sources and infection rate, the discrepancy should attribute to the underlying spreading mechanism.

We further examine the structural difference between real viral spreading and simulated epidemic diffusion. In Fig. \ref{SIR}c, we show the proportion of diffusion trees with a certain depth. The ratio between real cases and SIR model is displayed in the inset. Compared with SIR model results, about $90\%$ realistic viral spreading only last for one generation. On the contrary, nearly $50\%$ SIR diffusion trees have multiple-step cascades. For diffusion trees with a certain depth, we show the proportion of spreading links in these trees in Fig. \ref{SIR}d and the ratio between real viral spreading and SIR simulations in the inset.  While multiple-step cascades are responsible for the majority ($60\%$) of spreading links in SIR model, over $80\%$ spreading links are limited to one-step diffusion in real viral spreading. On the whole, SIR model overestimates the significance of multiple-step cascades and fails to reproduce the spreading pattern in real scenario. Therefore, our observation questions the previous adoption of epidemic-driven models in the research of information diffusion.

Since the social spreading pattern is coupled with other types of spreading patterns in the process of information diffusion, and the links in social spreading pattern may be affected by other spreading types even though the other links are deleted, the difference between social spreading pattern may come from the interaction among the spreading pattern. To make clear of this point, we eliminate the effect of the interaction among the spreading patterns and compare the real information diffusion with SIR model. We extract 231,333 diffusion trees entirely composed of social spreading pattern, which have no interaction with other diffusion types. The analysis results are shown in Fig. \ref{SIR}e and Fig. \ref{SIR}f. Clearly, there exists fundamental discrepancy between real diffusion and SIR model. Therefore, the difference should stem from the spreading mechanism, rather than the interaction among spreading patterns.

\begin{figure}[ht]
\begin{center}
\includegraphics[width=1\columnwidth]{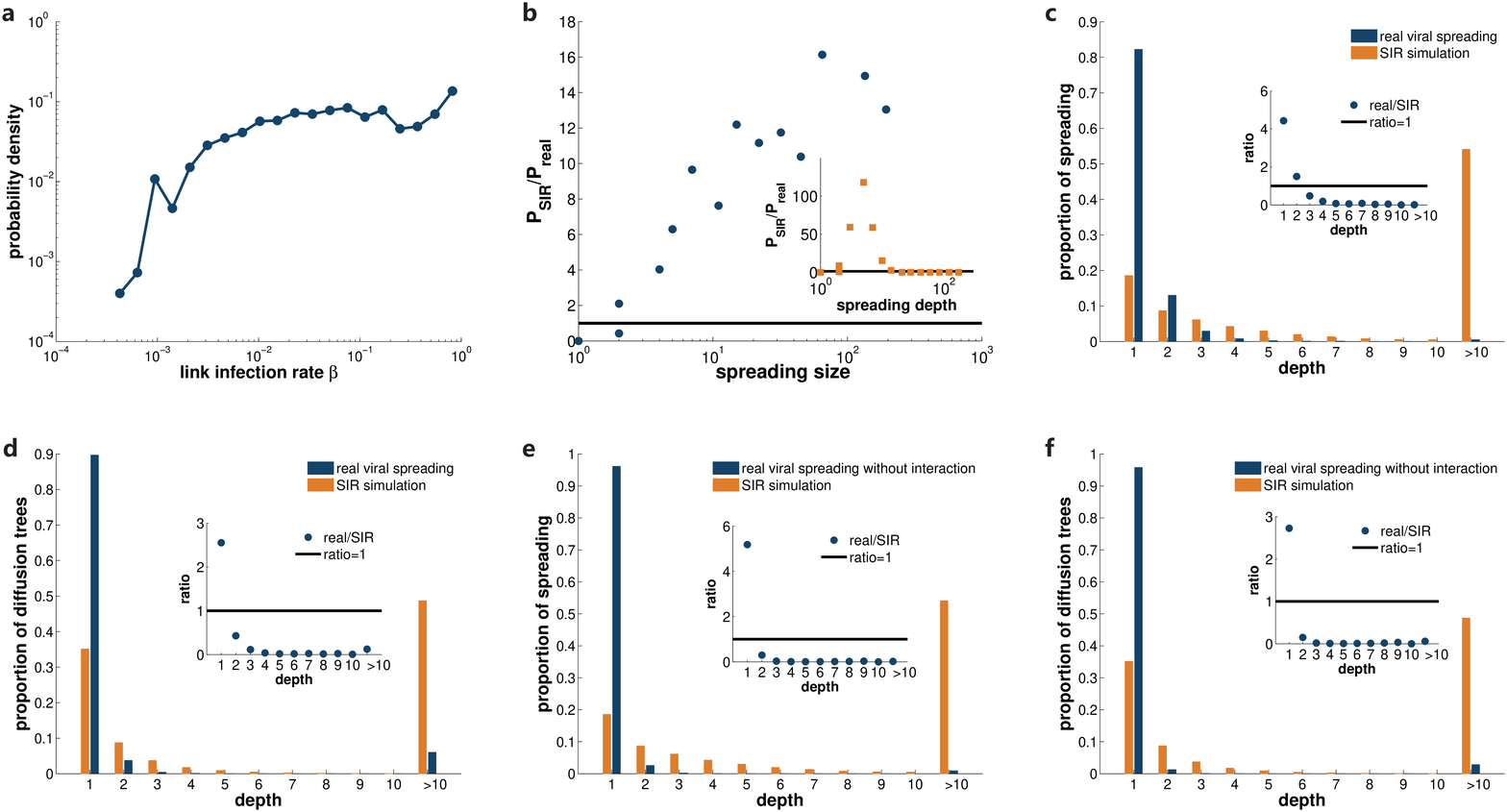}
\end{center}
\caption{{\bf SIR modeling with links' infection rate cannot reproduce the realistic viral spreading pattern.} {\bf a}, distribution of the real-world infection rate for each social link $\beta$ calculated from viral spreading instances. The ratio between the size distribution of SIR simulations and real viral spreading is displayed in {\bf b}. Inset shows the ratio of depth distribution. In {\bf c}, the proportion of diffusion trees with a given depth for both SIR simulations and real viral spreading is presented, and the ratio between real cases and SIR modeling is shown in the inset. {\bf d} illustrates the proportion of spreading instances for diffusion with a given depth for both cases. The inset shows the ratio between real viral spreading and simulations. Same analyses are shown in {\bf e} and {\bf f} for real viral spreading without interactions with self-promotion and broadcast diffusion. } \label{SIR_link}
\end{figure}

In the above experiment, we perform SIR model with individuals' infection rate defined by the fraction of neighbors who repost one's information. However, in general, each user should exert different influence on distinct neighbors. Considering this fact, we also conduct SIR simulation with infection rate inferred for each link. For a given directed social link, the infection rate is calculated as follows: we divide the number of diffusion instances from the source node to the end node by the total number of posts published by the source node in all social spreading trees. The distribution of links' infection rate is displayed in Fig. \ref{SIR_link}a, which is also heterogeneous. We perform same analysis for SIR model with links' infection rate and present the results in Fig. \ref{SIR_link}b-f. For both viral spreading with and without interaction with other diffusion types, SIR model with links' infection rate cannot reproduce the realistic viral spreading pattern - it also overestimates the importance of multiple-step cascades.

\subsection*{Self-promotion}

\begin{figure}[ht]
\begin{center}
\includegraphics[width=0.8\columnwidth]{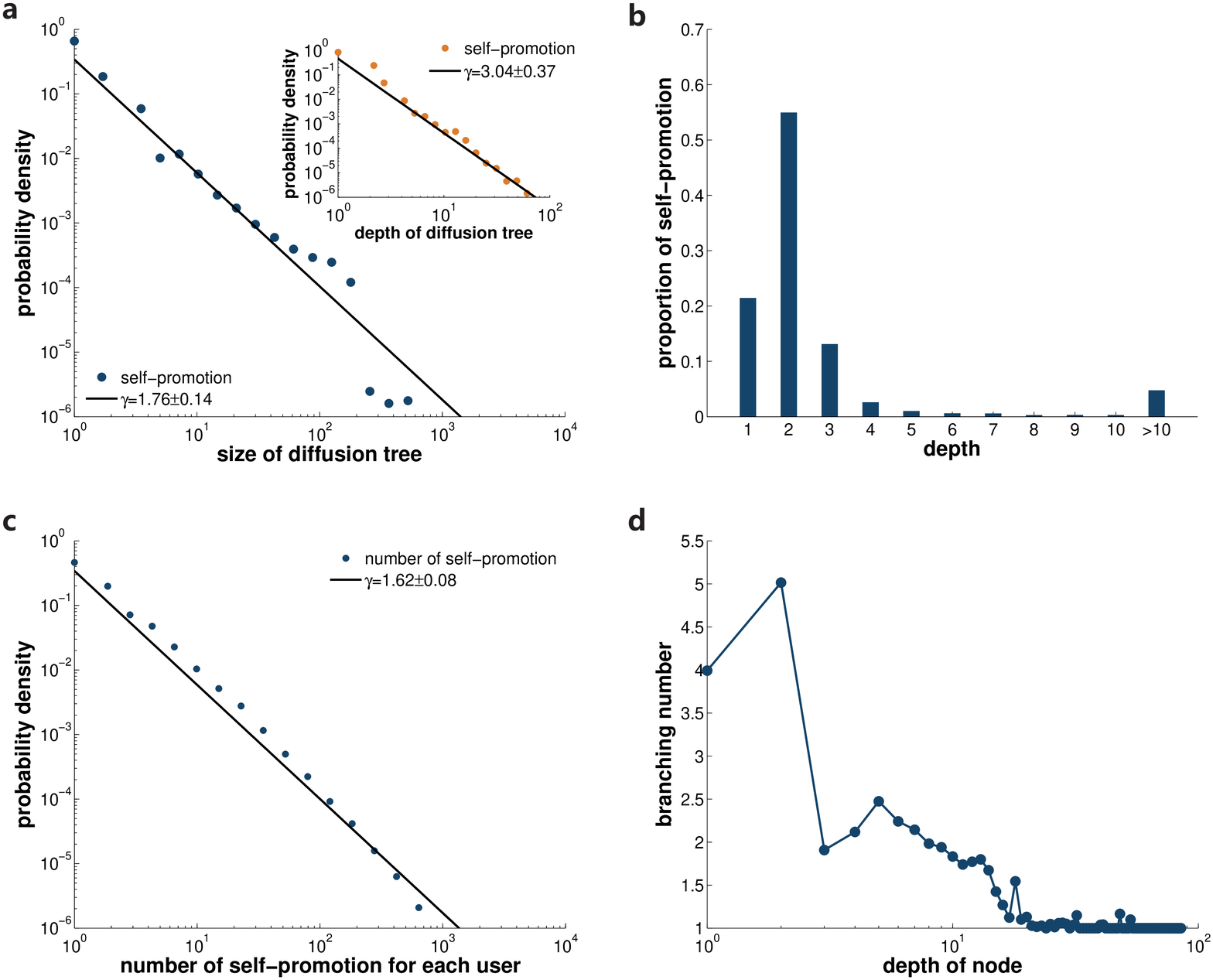}
\end{center}
\caption{{\bf Analysis of the self-promotion.} {\bf a} shows the distributions of the size and depth of self-promotion diffusion trees. The fraction of self-promotion links in diffusion trees with a certain depth is displayed in {\bf b}. In {\bf c} we present the probability distribution of the total number of self-promotion for each user, which has a power-law shape with exponent $\gamma=1.62\pm0.08$. In {\bf d} we plot the relationship between posts' branching number and their depth in self-promotion diffusion trees.} \label{selfpromotion}
\end{figure}

The diffusion type of self-promotion makes up a large fraction of the information diffusion. Self-promotion enhances the exposure of information, thus increasing the potential of being cited by other posts. Among the 315,937 individuals participating in information spreading, only 53,835 users have the behavior of self-promotion. We reconstruct the diffusion trees of self-promotion as before and obtain 254,861 diffusion trees. The distribution of the size and depth of these trees are displayed in Fig. \ref{selfpromotion}a. Both of them follow pow-law distributions approximately. However, compared with the diffusion trees of social spreading, the maximum size and depth are much smaller. For instance, the maximum depth of social spreading can be up to several hundreds, while the maximum depth is only 85 in case of self-promotion. Among all the self-promotion instances, $92.13\%$ are contributed by the diffusion trees that last less than 5 generations (see Fig. \ref{selfpromotion}b). Different from social spreading in which over $80\%$ spreading links locate in one-step diffusion, self-promotion trees with depth 2 contribute above $50\%$ of all spreading links. Intuitively, this phenomenon can be explained by the behavioral preference of each diffusion type. For social spreading, people usually place more trust in their direct neighbors. Therefore, most social spreading should concentrate in the first layer. On the contrary, if users decide to self-promote their own posts, they usually do not stop at the original ones. They would continue self-promote their subsequent posts, attempting to make more exposures. That may be the reason why most self-promotion resides in multi-step diffusion trees.

When we map the self-promotion trees to the diffusion among users, the process is simple since each self-promotion tree corresponds to a unique user. Figure \ref{selfpromotion}c shows that the total number of self-promotion for each user follows a power-law distribution with an exponent $\gamma=1.62\pm0.08$. This means while most of the users perform self-promotion for very few times, there exist a small number of users self-promote their posts frequently. If we plot the relationship between posts' branching number and their depth in the self-promotion diffusion trees, we find that while posts in the first few layers can have an either extremely large or small branching number, most of the posts located deep in the trees have branching number of 1. Therefore in case of self-promotion, users are more motivated to repost their earlier posts, which appears in the first few generations in the self-promotion diffusion trees.

\subsection*{Broadcast}

\begin{figure}[ht]
\begin{center}
\includegraphics[width=0.8\columnwidth]{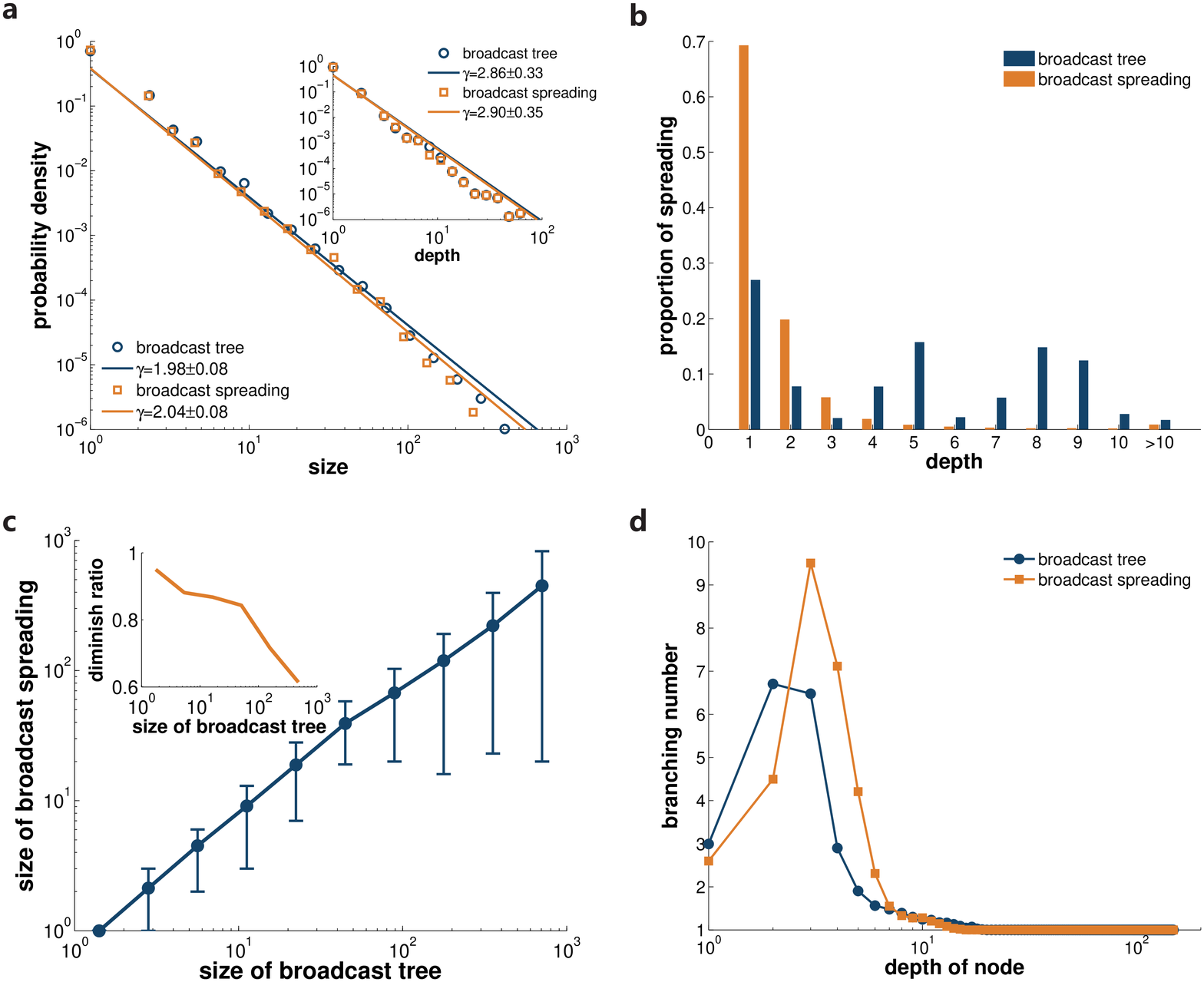}
\end{center}
\caption{{\bf Properties of the broadcast.} In {\bf a}, we display the distributions of the size and depth for broadcast diffusion trees and broadcast spreading respectively. The proportion of broadcast links in diffusion processes with a certain depth is shown in {\bf b}. The relation between the size of broadcast spreading and broadcast diffusion trees is displayed in {\bf c}. Error bars indicate $10\%$ and $90\%$ percentiles. The inset presents the diminishing ratio when mapping the diffusion trees to broadcast spreading. We plot nodes' average branching number versus their depth in diffusion in {\bf d}. } \label{broadcast}
\end{figure}

Similar with the analysis of social spreading and self-promotion, we reconstruct 441,027 diffusion trees of broadcast. As shown in Fig. \ref{broadcast}a, the distribution of the size and depth of these trees are power-law, with exponents $\gamma=1.98\pm0.08$ and $\gamma=2.86\pm0.33$ separately. Although the distribution is similar with that of social spreading and self-promotion, there exists a remarkable difference in the constitution of broadcast diffusion trees. From Fig. \ref{broadcast}b, we can conclude that a considerable fraction of broadcast occurs in diffusion trees last for many generations, which opposes to the cases of social spreading and self-promotion. Considering the power-law distribution of tree depth, our observation indicates that even though the deep diffusion trees are rare, their scales are extremely large so that the broadcast links in these trees can occupy a significant fraction of the total broadcast instances. This structural difference may highlight the crucial role of broadcast playing in the information diffusion in LJ community. In Fig. \ref{broadcast}d, the relationship between the branching number and nodes' depth also holds for broadcast, i.e. the branching number for nodes in less than 20 layers varies in a wide range and has mean values larger than 1, whereas nodes deeper than 20 generations have only one subsequent broadcast.

Since one user can perform broadcast for many times, the population participating in a broadcast diffusion tree should be smaller than the tree size. We denote the spreading processes among users in broadcast as broadcast spreading. In Fig. \ref{broadcast}c, we display the relationship between the size of broadcast spreading and corresponding diffusion trees. Compared with the case of social spreading, the diminish ratio is higher for deep diffusion trees. This means in broadcast the diffusion trees are capable of reaching larger population under the same circumstances. The size and depth of broadcast spreading are also power-law distributed, with exponents $\gamma=2.04\pm0.08$ and $\gamma=2.90\pm0.35$ respectively(see Fig. \ref{broadcast}a). In broadcast spreading, $94.91\%$ of the involved users repost information for the first time in less than 4 steps, as displayed in Fig. \ref{broadcast}b. Recall that for broadcast diffusion trees, deep ones are responsible for a large fraction of broadcast, our finding implies that after users have reposted the information for the first time, they still keep citing the information in other posts so that the diffusion trees can grow deeper. Same as previous analysis, we plot the branching number of nodes versus their depth in the broadcast spreading in Fig. \ref{broadcast}d. Deep nodes only lead to one subsequent propagation. Most of the users adopt the information in just a few generations.

\subsection*{Human activity}

\begin{figure}[ht]
\begin{center}
\includegraphics[width=1\columnwidth]{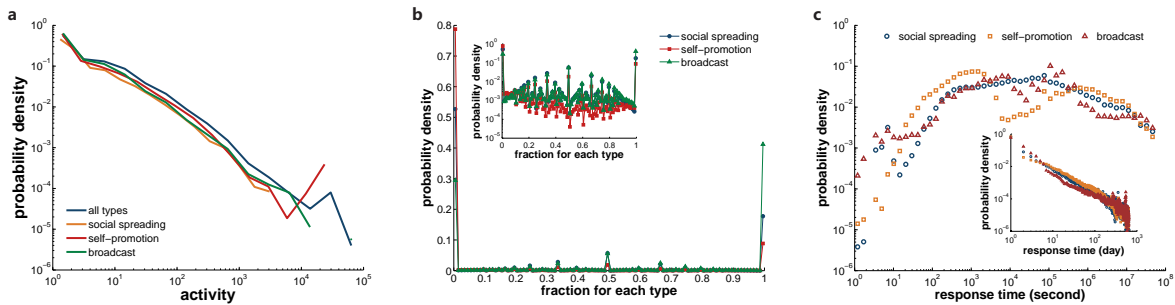}
\end{center}
\caption{{\bf The human activity of LJ users.} In {\bf a} we show the distribution of users' activity for each mechanism. For each user, we calculate the fraction of each mechanism that the selected user has conducted, and present the distribution of obtained fraction in {\bf b}. In the inset we change the linear scale of y axis to a logarithmic scale. {\bf c} displays the distribution of response time $\tau$ for social spreading, self-promotion, and broadcast. We adopt the time unit of second in the main panel and day in the inset.} \label{activity}
\end{figure}

One critical factor affecting the spreading outcome is the human activity, including activity frequency \cite{Muchnik2013} and response time \cite{Iribarren2009}. To be precise, the activity frequency is defined as the number of each user's participation in a specific diffusion type, such as social spreading, self-promotion and broadcast. For example, if a user is observed to have cited $n$ posts of his/her neighbors, his/her activity frequency of social spreading is defined as $n$. In addition, the response time is defined as the time it takes for an individual to repost the information. For instance, if at time $t_1$ a user cited a post which was published at time $t_0$, the response time of this diffusion instance is $t_1-t_0$. In Fig. \ref{activity}a, we display the distributions of users' activity number of social spreading, self-promotion, broadcast as well as the overall diffusion. They are extremely similar except in the range of tails. The maximum number of social spreading is much smaller than that of self-promotion and broadcast, which may be due to the existence of robot users who automatically repost their own publications or gather other users' popular posts.

One may wonder if users tend to conduct all types of diffusion or just incline to adopt only one type. In order to answer this question, for each user, we calculate the fraction of each diffusion mechanism. For instance, if a user performs $a_{1}$ social spreading, $a_{2}$ self-promotion, $a_{3}$ broadcast and a total of $A=a_{1}+a_{2}+a_{3}$ diffusion, then the fraction for these three types are $a_{1}/A$, $a_{2}/A$, and $a_{3}/A$. We show the distribution of the obtained fraction for each type in Fig. \ref{activity}b. A shocking phenomenon is that for all the cases, the fraction concentrates near the values of $0$ and $1$, while the intermediate values are relatively rare. This implies the spreading behavior of users is highly personalized: most of the users choose to adopt only one specific spreading behavior. Therefore, our classification of spreading types is related to users' inherent behaviors. Apart from users preferring to repost their peers' posts, there exist considerable numbers of users who only adopt self-promotion and broadcast. The function of these users in information spreading will be discussed in the next section.

Another important factor of human activity is the response time $\tau$ \cite{Barabasi2005}. In Fig. \ref{activity}c, we plot the distribution of response time for each diffusion type, adopting the time unit of second in the main panel and day in the inset. The response time ranges over several magnitude, which exhibits extremely heterogeneity. Vast majority of spreading takes place within the time interval between 0.1 to 10 days. Therefore the characteristic time of information spreading in LJ is about one day, which is far smaller than the time scale of the formation of network structure. If we change the time unit from second to day, all three types of diffusion follow a similar power-law distribution. Previous research has shown that the power-law distribution of human activity can have a significant impact on information spreading \cite{Iribarren2009}.

\subsection*{Dynamical coupling of different spreading mechanisms}

\begin{figure}[ht]
\begin{center}
\includegraphics[width=1\columnwidth]{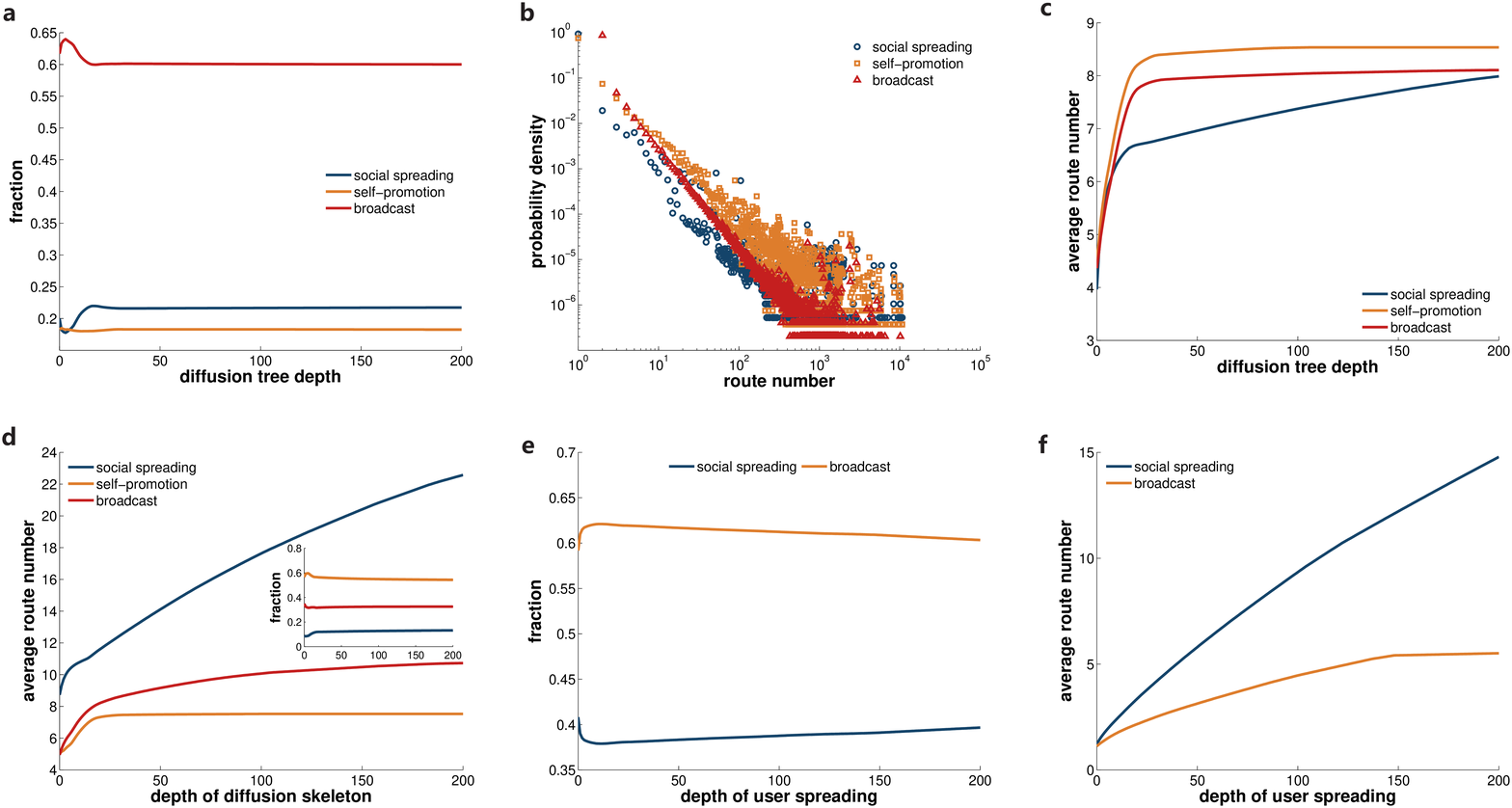}
\end{center}
\caption{{\bf Coupling of distinct mechanisms.} We plot the fraction of the social spreading, self-promotion, and broadcast links in diffusion trees deeper than a given threshold in {\bf a}. The x-axis value is the lower bound of selected trees' depth. In {\bf b}, the distribution of the route number for each type is displayed. We calculate the average route number of the diffusion links for each type in diffusion trees deeper than a certain depth, and present the results in {\bf c}. After removing the leaves of diffusion trees, we obtain the information diffusion skeletons. We show the average route number and composition in diffusion skeletons whose depth exceeds certain values in the main panel and inset of {\bf d} respectively. In the spreading processes among population, the fraction and average route number of social spreading and broadcast links are presented in {\bf e} and {\bf f}.} \label{rnandfrac}
\end{figure}

Since the diffusion trees of posts are induced by the combination of distinct mechanisms, it is desirable to investigate how these spreading types couple with each other. First we should check the fraction of each type in the diffusion trees. On the whole, the diffusion trees are composed of $20\%$ social spreading, $18.3\%$ self-promotion, and $61.7\%$ broadcast. Note that this composition is different from that of diffusion links, which is  $26.8\%$, $31.14\%$, and $42.06\%$. The reason for this discrepancy lies in that posts may contain several hyperlinks to other posts. As a consequence, a single spreading instance could participate in more than one diffusion trees. On average, broadcast posts contain more referred hyperlinks, which raises the proportion of broadcast. In order to check the composition of diffusion trees with different depths, we present the proportion of the social spreading, self-promotion, and broadcast links in diffusion trees deeper than a given threshold in Fig. \ref{rnandfrac}a. We first select diffusion trees whose depth exceeds a given value, and then calculate the fraction of each mechanism in these trees. As the lower bound of depth increases, the fraction becomes steady. This indicates that the composition of diffusion trees is similar for each depth.

To quantify the importance of spreading links in diffusion trees, we define the route number $r(i,j)$ for link $(i,j)$ as the number of shortest paths from the information source to the other nodes passing through this link:
\begin{equation}
r(i,j)=\sum_{t\in V\setminus \{s\}}\sigma_{st}(i,j), \label{routenumber}
\end{equation}
where $V$ is the set of nodes in a diffusion tree and $\sigma_{st}(i,j)$ is the number of shortest paths between the source node $s$ and another node $t$ which pass through link $(i,j)$. Apparently, $r(i,j)$ stands for the number of subsequent diffusion induced by link $(i,j)$. The distributions of route number for all the types are displayed in Fig. \ref{rnandfrac}b, which are all highly skewed. In Fig. \ref{rnandfrac}c, with the increase of the lower bound of selected trees' depth, the average route number $\langle r\rangle$ of each mechanism first grows rapidly and then becomes steady, except that $\langle r\rangle$ of social spreading still grows at a slower pace. Although self-promotion is fewer than social spreading and broadcast, the average route number is larger, which reflects the significant role of self-promotion in the formation of diffusion trees.

Aimed to explore the formation of diffusion trees, we remove the leaves of diffusion trees and check the composition of the obtained {\it diffusion skeletons}. We focus on this structure since the diffusion trees are expanded based on the diffusion skeletons. In fact, $85.24\%$ nodes located in the leaves are induced by the other $14.76\%$ nodes in diffusion skeletons. Therefore these skeletons determine the top-level structure of diffusion trees. In the inset of Fig. \ref{rnandfrac}d, we can see self-promotion is responsible for up to $60\%$ links in diffusion skeletons. This implies that large numbers of social spreading and broadcast links are in the leaves of diffusion trees, thus pulling down their average route numbers. This would lead to the higher average route number of self-promotion in Fig. \ref{rnandfrac}c . If we check the average route number in diffusion skeletons, surprisingly, social spreading has a much higher value (see Fig. \ref{rnandfrac}d). This indicates that social spreading is crucial in the formation of diffusion skeletons. Although the number of social spreading links is smaller, they incline to locate near the information source, which could induce more subsequent links in skeletons.

As we have pointed out, larger diffusion trees may not necessarily lead to more influenced population. Considering in practice people care more about the number of influenced individuals, we need to explore the dissemination among users. In the following analysis, we map the diffusion trees to spreading among users as we did for social spreading and broadcast. The obtained user spreading only contains links of social spreading and broadcast, since self-promotion links are not capable of producing new adopters. As can be seen in Fig. \ref{rnandfrac}e, broadcast links still hold the majority of user spreading, regardless of our selection of the lower bound for user spreading's depth. The average route number of social spreading is higher than that of broadcast in Fig. \ref{rnandfrac}f. This indicates that, despite that broadcast links' number is higher, social spreading could on average lead to more adopters. This observation highlights the function of social spreading in expanding influenced population.

\begin{figure}[ht]
\begin{center}
\includegraphics[width=0.8\columnwidth]{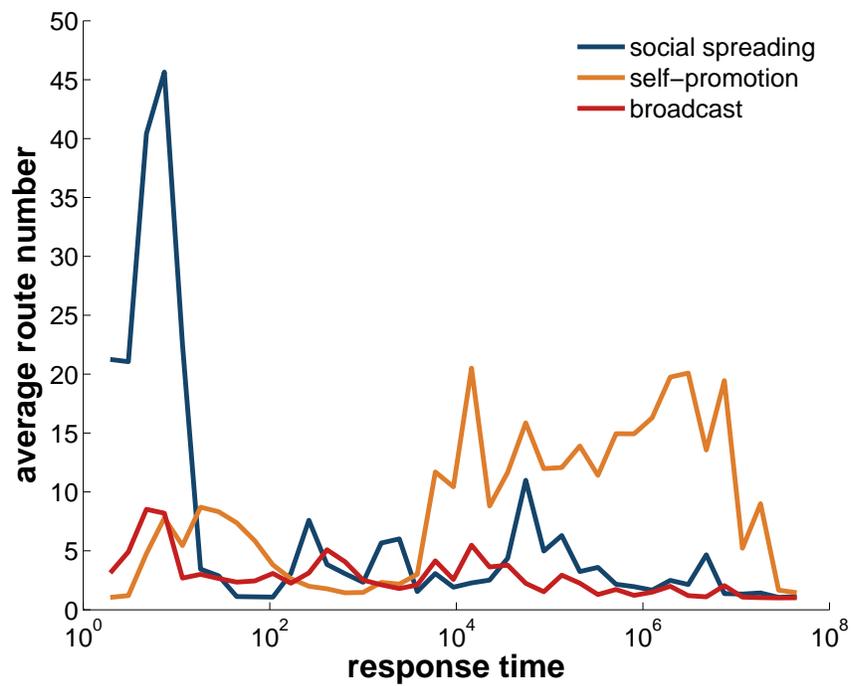}
\end{center}
\caption{{\bf The average route number versus response time for different mechanisms.} For each diffusion type, we classify diffusion links according to their response time (in second), or equivalently the diffusion speed, and display the average route number for each case. } \label{rnvstime}
\end{figure}

One may wonder if the route number is related to the spreading speed. Therefore we display the average route number versus response time for three spreading patterns in Fig. \ref{rnvstime}.  For social spreading, it is striking that fast spreading instances with small response time have extremely large route numbers. This implies, on average, the faster one reposts his/her neighbors' information, the more subsequent diffusion he/she will induce. While for self-promotion, it is slow spreading that has larger route number. For broadcast, the average route number has no clear relationship with spreading speed, except a downward trend with growing response time. How these differences arise is an interesting topic to be further explored.

\section*{Discussion}

Understanding human behaviors associated with dissemination of information and the resulting information propagation patterns in online social communities is crucial for a wide range of applications. Despite the vast and growing literature on information spreading, the relationship between diffusion and users' dissemination patterns has not been explored. Here, we perform a detailed analysis on the diffusion data and social network structure of an online blog community. To our surprise, we find that most users exhibit persistent behavior following one of the three patterns - social spreading, self-promotion and broadcast. We study diffusion trees of each type and show that majority of cases of information propagation are limited to the first few generations. In particular, we compare the spreading pattern of real information diffusion and SIR model through simulations with realistic infection rate. The discrepancy in spreading pattern indicates that widely used epidemic models are incapable of reproducing realistic information spreading, which necessitates more accurate information spread models. Moreover, the claim about the prominence of social spreading in information dissemination is also supported by the fact that social spreading can lead to more subsequent individuals acquiring the information.

The suggested classification of the information diffusion patterns can in future research be applied to other online social networks, including Facebook and Twitter. Further research is necessary to understand how each of the mechanisms associates with user traits, content and the outcomes of the information diffusion. We believe, that our classification can be employed in new generation of information spread models and for practical use, such as marketing to control information flow in social networks.

% Do NOT remove this, even if you are not including acknowledgments
\section*{Acknowledgments}

%We are grateful for the suggestions from anonymous reviewers. SP, ST and ZZ are supported by Major Program of National Natural Science Foundation of China (No. 11290141), NSFC (No. 11201018), International Cooperation Project No. 2010DFR00700, Fundamental Research of Civil Aircraft No. MJ-F-2012-04. SP also acknowledges support from Innovation Foundation of BUAA for PhD Graduates.

%\section*{References}
% The bibtex filename
\bibliography{template}

\section*{Figures}

%\section*{Tables}
%\begin{table}[!ht]
%\caption{
%\bf{Table title}}
%\begin{tabular}{|c|c|c|}
%table information
%\end{tabular}
%\begin{flushleft}Table caption
%\end{flushleft}
%\label{tab:label}
% \end{table}

\end{document}